\documentclass[a4paper,fleqn]{cas-dc}
\usepackage{soul}
\usepackage[sort&compress,numbers]{natbib}
\usepackage{lineno}
\usepackage{setspace}
\usepackage{multirow}

\usepackage{cuted}

\def\tsc#1{\csdef{#1}{\textsc{\lowercase{#1}}\xspace}}
\tsc{WGM}
\tsc{QE}
\tsc{EP}
\tsc{PMS}
\tsc{BEC}
\tsc{DE}

\begin{document}
\let\WriteBookmarks\relax
\def\floatpagepagefraction{1}
\def\textpagefraction{.001}
\shorttitle{Charge Transport Mechanism in Chevron--Graphene Nanoribbons}
\shortauthors{Pereira J\'unior \textit{et~al}.}

\title [mode = title]{Charge Transport Mechanism in Chevron--Graphene Nanoribbons}

\author[1]{Marcelo Lopes Pereira J\'{u}nior}
\author[1]{Wiliam Ferreira da Cunha}
\author[2]{Rafael Tim\'oteo de Sousa Junior}
\author[2]{William Ferreira Giozza}
\author[1]{Geraldo Magela e Silva}
\author[1,3]{Luiz Antonio Ribeiro J\'{u}nior}
\cormark[1]
\ead{ribeirojr@unb.br}

\address[1]{Institute of Physics, University of Bras\'ilia, Bras\'ilia, Brazil.}
\address[2]{Department of Electrical Engineering, University of Bras\'{i}lia 70919-970, Brazil.}
\address[3]{PPGCIMA, Campus Planaltina, University of Bras\'{i}lia, 73345-010, Bras\'{i}lia, Brazil.}

\cortext[cor1]{Corresponding author}

\begin{abstract}
	From the moment atomic precision control of the growth process of graphene was achieved, more elaborated carbon allotropes were proposed opening new channels for flat optoelectronics at the nanoscale. A special type of this material presenting a V-shape (or "kinked" pattern) was recently synthesized and named Chevron-graphene nanoribbons (C--GNRs). To realize the reach of C--GNRs in developing new applications, the formation, and transport of charge carriers in their lattices should be primarily understood. Here, we investigate the static and dynamical properties of quasiparticles in C--GNRs. We study the effects of electron-phonon coupling and doping on the system. We also determine the kind of charge carriers present in C--GNR. It is observed that a phase transition occurs between a delocalized regime of conduction and regimes mediated by charge carriers. Such a phase transition is highly dependent on the doping concentration. Remarkably, crucial differences from the transport in standard graphene nanoribbons are identified. These factors are noted to have a profound impact on the mobility on the system which, in turn, should decisively impact the performance of electronic devices based on C--GNRs. 
\end{abstract}




\maketitle
\doublespacing

\section{Introduction}
Molecular electronics is currently a field attracting the attention of the scientific community for the potential it presents of giving rise to low-cost energy-efficient devices. Organic systems, in particular, are the most promising source of materials to spawn a molecular electronic revolution. This feature is due to the cost-efficient nature of the material as well as to the similarities between Carbon and Silicon, over which the current electronic industry is based.

Among the several possible organic systems to lead the electronic industry, the synthesis of graphene quickly defined it as the most promising one \cite{Novoselov666,geim}. Being a true two-dimensional system it provides a great surface to volume ratio for several applications. Also, the possibility of roll-to-toll processing \cite{roll} resulted in the proposal of a large range of ideas \cite{ming}. As a major drawback, though, the pristine two-dimensional graphene sheet lacks the bandgap typical of the conventional semiconductors \cite{Novoselov,Zhou}.   
Several different procedures have been employed to circumvent this difficulty. The main one is to resort to graphene-based systems with different topological structures. Different allotropes have been extensively studied, among which are carbon nanotubes \cite{PhysRevB.96.075133} and nanoribbons \cite{doi.org/10.1038/nature07919}. The change in physical structure results in changes in the band structure of the materials, which reflects on the arising of interesting --- and frequently tunable --- electronic and transport properties.

\begin{figure*}[]
	\centering
	\includegraphics[width = 0.23\linewidth, angle = 90]{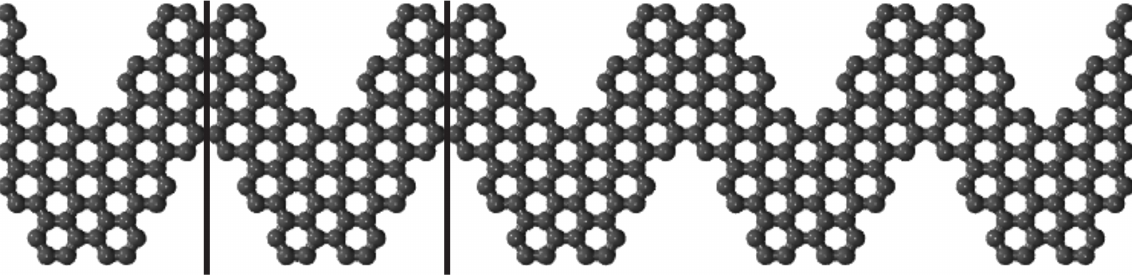}
	\caption{Structural representation of a chevron--graphene nanoribbon. An unit cell is highlighted.}
	\label{fig1}
\end{figure*}

The precision control of width, edge structure, stacking, and the growth process leads to more elaborated classes of graphene-based systems, opening a whole new range of possibilities. Carbon nanoscrolls  \cite{doi:10.1021/nl900677y}, popgraphene \cite{wang2018popgraphene}, phagraphene \cite{wang2015phagraphene}, pentagraphene \cite{zhang2015penta}, and Chevron-graphene \cite{chev} nanoribbons (C--GNRs) are among the new candidates. In recent years, this latter species has been the subject of major interest. It consists of a special type of graphene nanoribbon that presents a V-shape, or "kinked" pattern. It can be reasoned to be formed from the periodic juxtaposition of two angularly shifted armchair graphene nanoribbons, displaced as shown in Figure \ref{fig1}. 

Recent experimental evidence suggests C--GNRs as a promising material for applications in organic electronics and photovoltaics \cite{C4CC00885E}. Several other works have dealt with the synthesis of C--GNR \cite{naturesynth,C4CC00885E,vosynth}. Some studies have shown that C-GNRs share interesting properties that arise in different carbon allotropes while presenting other unique features \cite{PMID:25779989,C4FD00131A}. Finally, possible applications of C--GNRs in heterojunctions have also already been reported \cite{xyzcaihet}. However, so far the literature lacks the theoretical description of its electronic and transport properties. This fact results in controversies in elementary, yet fundamental, characteristics of the system. Therefore, an extensive study describing how electronic transport takes place in C--GNRs is paramount to the further development of the molecular electronics. Of particular importance is to highlight the differences between transport phenomena in C--GNRs and those observed in regular armchair graphene nanoribbons.   

Herein we make use of a 2D tight-binding Hamiltonian, which includes lattice relaxation effects to discuss the electronic and transport properties in C--GNRs. We start by conducting a set of simulations to determine the electron-phonon coupling strength. Following, we analyze how different electron-phonon coupling constants contribute to the appearance of quasiparticles. We discuss both the static and dynamical properties of different quasiparticles under the effect of an external electric field. This description is expected to give important insights to improve the performance of electronic devices based in C--GNRs, such as heterojunctions. 

\section{Methodology}
The model implemented to investigate the electronic and transport properties of C--GNRs is based on a 2D SSH-type Hamiltonian \cite{su_PRB,su_PRL} endowed with lattice relaxation. In it, we describe the electronic degrees of freedom of the system quantum mechanically in a second quantization formalism whereas the lattice is treated classically employing a set of Euler-Lagrange equations. It should be noted, however, that these two realms are coupled so that the problem ought to be solved self consistently.

The coupling between electronic and lattice degrees of freedom is implemented in the electronic transfer integrals, which are expressed in such a way that the hopping of electrons between sites depends on the lattice disposition. In the case of covalent C---C bonds, the lattice displacement is a small fraction of the equilibrium size (usually not much higher than 2\%). This justifies the use of a first-order expansion for the electronic transfer integral of $\pi$ electrons dependence on the displacement \cite{da2016polaron}.

\begin{figure}[]
	\centering
	\includegraphics[width = 0.85\linewidth]{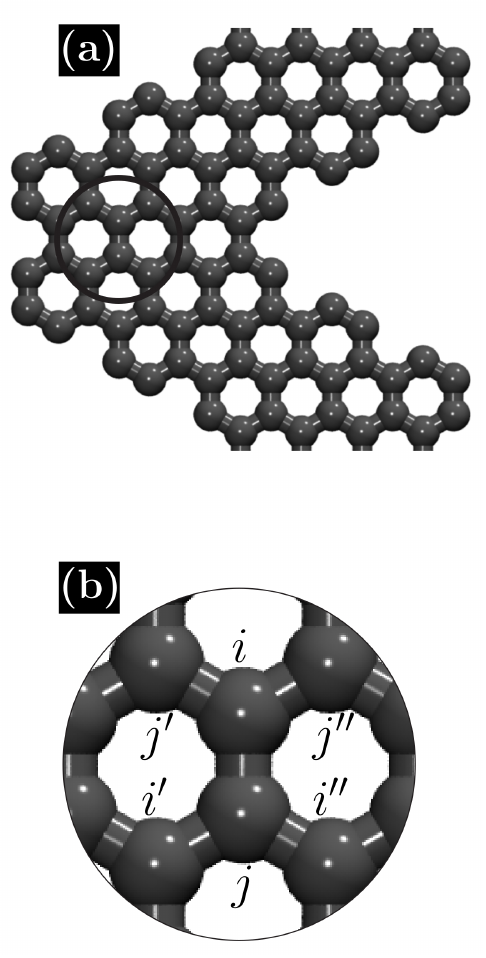}
	\caption{(a) Unit cell of a Chevron--Graphene Nanoribbon and (b) highlight presenting site labelling.}
	\label{fig2}
\end{figure}

Figure \ref{fig2}(a) shows the unit cell of a C--GNR lattice. The zoomed panel (Figure \ref{fig2}(a)) clarifies the labeling of the sites. Let $\eta_{i,j}$ be the variations in the bond-lengths between two neighboring sites $i$ and $j$. From the previous considerations, the expanded transfer integral to be plugged into our model Hamiltonian takes the form: 
\begin{equation}
	\label{hop}
	t_{i,j} = t_0 - \alpha \eta_{i,j}.
\end{equation}
$ t_0 $ is the hopping integral that the evenly spaced system would have in a pure tight binding model. $ \alpha $ is the aforementioned coupling constant, responsible for the coupling between electrons and phonons. 

Within the this approach, the model Hamiltonian of the system is given by
\begin{equation}
	\begin{split}
		H &= -\sum_{\langle i,j \rangle, s} \left(t_{i,j}^{\phantom{*}} C_{i,s}^\dag C_{j,s}^{\phantom{\dag}}+t_{i,j}^* C_{j,s}^\dag C_{i,s}^{\phantom{\dag}} \right ) \\
		&\hspace{0.1cm}+ \frac12K\sum_{\langle i,j \rangle} \eta_{i,j}^2+\frac1{2M}\sum_ip_i^2.
	\end{split}
\end{equation}
Here $ \langle i,j \rangle $ represents the sum carried out on neighboring sites \cite{su_PRB,su_PRL}. $C_{i,s}^{\phantom{\dag}}$ stands for the $\pi$-electron annihilation operator on site $i$ with spin $s$ and $ C_{i,s}^{\dag} $ represents its hermitian conjugate, i.e., the corresponding creation operator. The remaining terms express the lattice degrees of freedom of the system. The second term is a harmonic approximation to model the effective potential associated with $\sigma$-bonds between carbon atoms, $K$ being the elastic constant. The last term expresses the kinetic energy of the sites in terms of their momenta $ p_i $ and mass $M$ of the carbon cores. One should note that, as the transfer integral is position-dependent, the movement of the sites indeed impacts the behavior of the electrons. This reason dictates that the lattice and the electron parts of the problem must be solved simultaneously in a self-consistent fashion, as previously argued.

Following previous theoretical and experimental works \cite{barone2006electronic,de2012electron,ribeiro2015transport, kotov2012electron,yan2007electric,neto2009c,yan2007raman,cunha_PRB,cunha_CARBON,fischer_CARBON,silva_SR}, the values of the model parameters were set to be: 2.7 eV for $t_0$ and 21 eV/\AA$^2$ for $K$. As for the $\alpha$, the nature of the present contribution dictates that we should perform a specific search of the suitable values by considering the range from  0.1 to 6.0 eV/\AA. The most suitable values are obtained using a tuning procedure that takes into consideration the known properties of the material. 

Starting the iteration from an initial set of coordinates $ \{\eta_{i,j}\} $, a self-consistent stationary solution (with $ p_i = 0 $) of the system is determined \cite{lima2006dynamical}. The ground state is obtained with the diagonalization of the electronic Hamiltonian. The result of the diagonalization procedure returns, as eigenvalues, the energies of the electronic states and, as eigenstates, the wave functions for the ground state. 

As for the lattice part of the problem, we take this initial states and the according terms on the Hamiltonian to obtain the expectation value of the system's Lagrangean, $ \langle L \rangle = \langle \Psi | L | \Psi \rangle $, where $ | \Psi \rangle $ is the Slater determinant. We, thus, have:
\begin{equation}
	\begin{split}
		\langle L \rangle & = \frac{M}{2}\sum_i \dot\xi_i^2-\frac12K\sum_{\langle i,j \rangle}\eta_{i,j}^2  \\
		& + \sum_{\langle i,j \rangle, s} \left( t_0-\alpha \eta_{i,j} \right)\left(B_{i,j}+B^*_{i,j}\right);
	\end{split}
\end{equation}
where,
\begin{equation}
	B_{i,j} \equiv \sum_{k,s}{'}\psi^*_{k,s}(i) \psi^{\phantom{*}}_{k,s}(j)
\end{equation}
is the term that connects the electronic and the lattice part of the system. The prime in the later summation denotes that the sum must be carried out over the occupied states that compound the Slater determinant. 

With such an expression, one can solve the Euler-Lagrange equation to obtain an initial set of coordinates for the positions of the atoms $\{ \eta_{i,j} \}$. From it we begin an auto-consistent procedure, where a corresponding electronic set $\{ \psi_{k,s}(i) \}$ is obtained. This gives rise to a new Lagrangean whose Euler-Lagrange equations, when solved, return yet a new set of coordinates $\{ \eta_{i,j} \}$. The idea of this self-consistent process is to iteratively repeat itself until a given convergence criterion is met. After convergence, one obtains a stationary solution $\{ \eta_{i,j} \}$ and $\{\psi_{k,s}(i)\}$. 

\begin{figure}[]
	\centering
	\includegraphics[width = \linewidth]{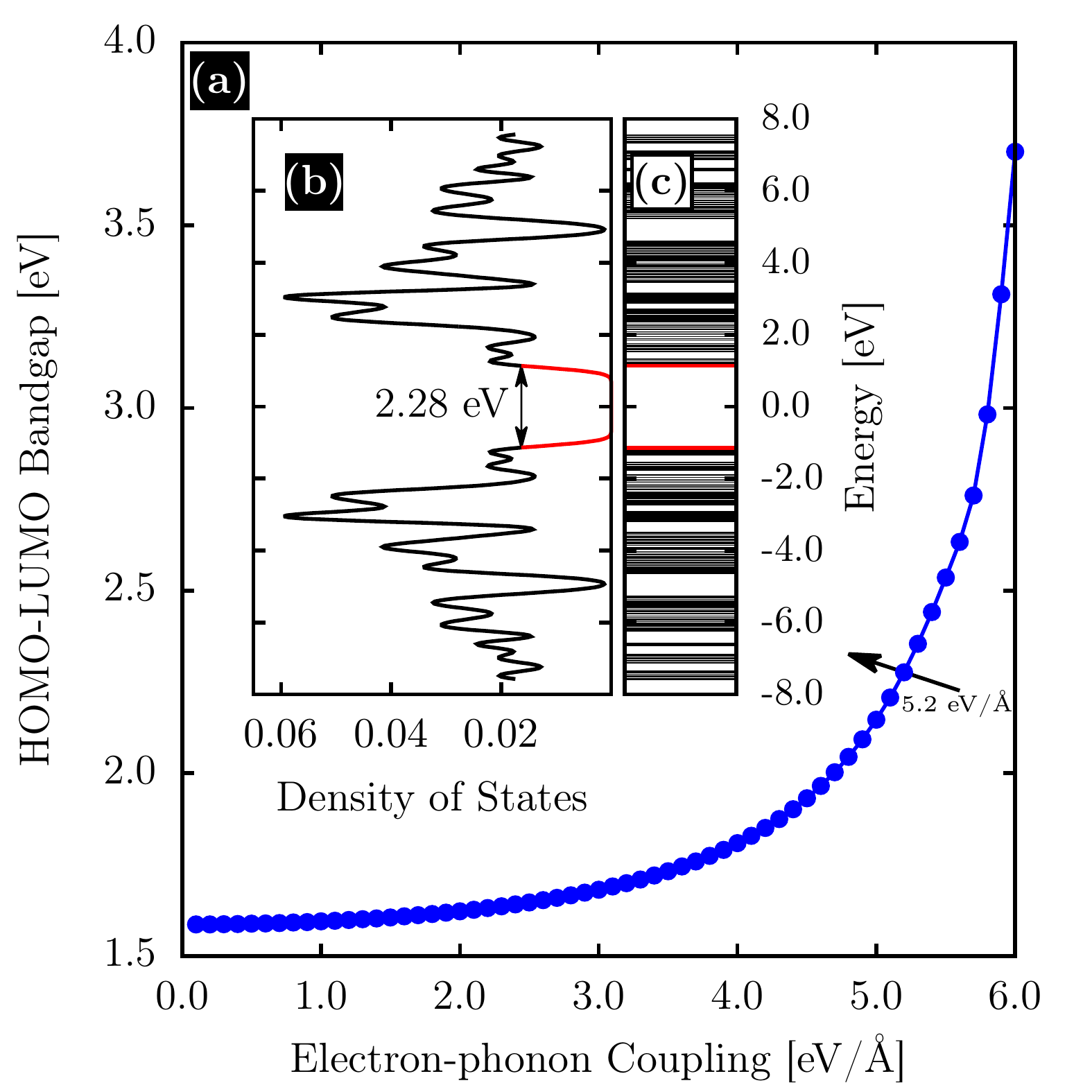}
	\caption{(a) HOMO-LUMO dependency on $\alpha$. (b) Density of States and (c) band structure of the system for $\alpha = 5.2 $ eV/\AA.}
	\label{fig3}
\end{figure}

In order to time evolve the system, we make use of the time-dependent Schr\"{o}dinger equation for electrons and obtain: 
\begin{equation}
	|\psi_{k,s}(t+dt)\rangle = e^{-\frac i\hbar H(t) dt}|\psi_{k,s}(t)\rangle.
\end{equation}
We thus expand $ |\psi_{k,s}(t)\rangle $ on a basis of eigenstates of the electronic Hamiltonian and perform numerical integration resorting to the same scheme reported in previous works \cite{lima2006dynamical,cunha_PRB}.

For the classical treatment of the lattice part of the system, the solution of the Euler-Lagrange equations is written as a Newtonian equation that describes the movement of the sites
\begin{equation}
	\begin{split}
		M\ddot \eta_{i,j} &= \frac12K\left(\eta_{i,j'}+\eta_{i,j''}+\eta_{j,i'}+\eta_{j,i''} \right )-2K\eta_{i,j}\\ 
		&+ \frac12\alpha\left(B_{i,j'}+B_{i,j''}+B_{j,i'}+B_{j,i''}-4B_{i,j} + \mathrm{c.c.}\right ).
	\end{split}
\end{equation}

Finally, we include an external electric field $\mathrm{\textbf{E}}(t)$ through a Peierls substitution on the electronic transfer integrals of the system, making the hopping term
\begin{equation}
	t_{i,j} = e^{-i\gamma\mathrm{A(t)}}\left(t_0 - \alpha \eta_{i,j} \right ).
\end{equation}
Here, $\mathrm{A}(t)=\mathrm{\textbf{A}}(t)\cdot \hat{\eta}_{i,j}$, $\mathrm{\textbf{A}}(t)$ being the vector potential whose relation to the electric field $\mathrm{\textbf{E}}(t)$ is given by $ \mathrm{\textbf{E}}(t) = -(1/c)\dot{\mathrm{\textbf{A}}}(t) $. $ \gamma \equiv ea/(\hbar c) $, with $a$ being the lattice parameter ($ a = 1.42 $ \AA~ in graphene nanoribbons), $e$ being the absolute value of the electronic charge, and $c$ the speed of light. In order to mitigate artificial numerical effects that tend to arise from the abrupt implementation of electric field in this kind of system, we turned it on adiabatically~\cite{da2016polaron}.

\begin{figure}[]
	\centering
	\includegraphics[width = \linewidth]{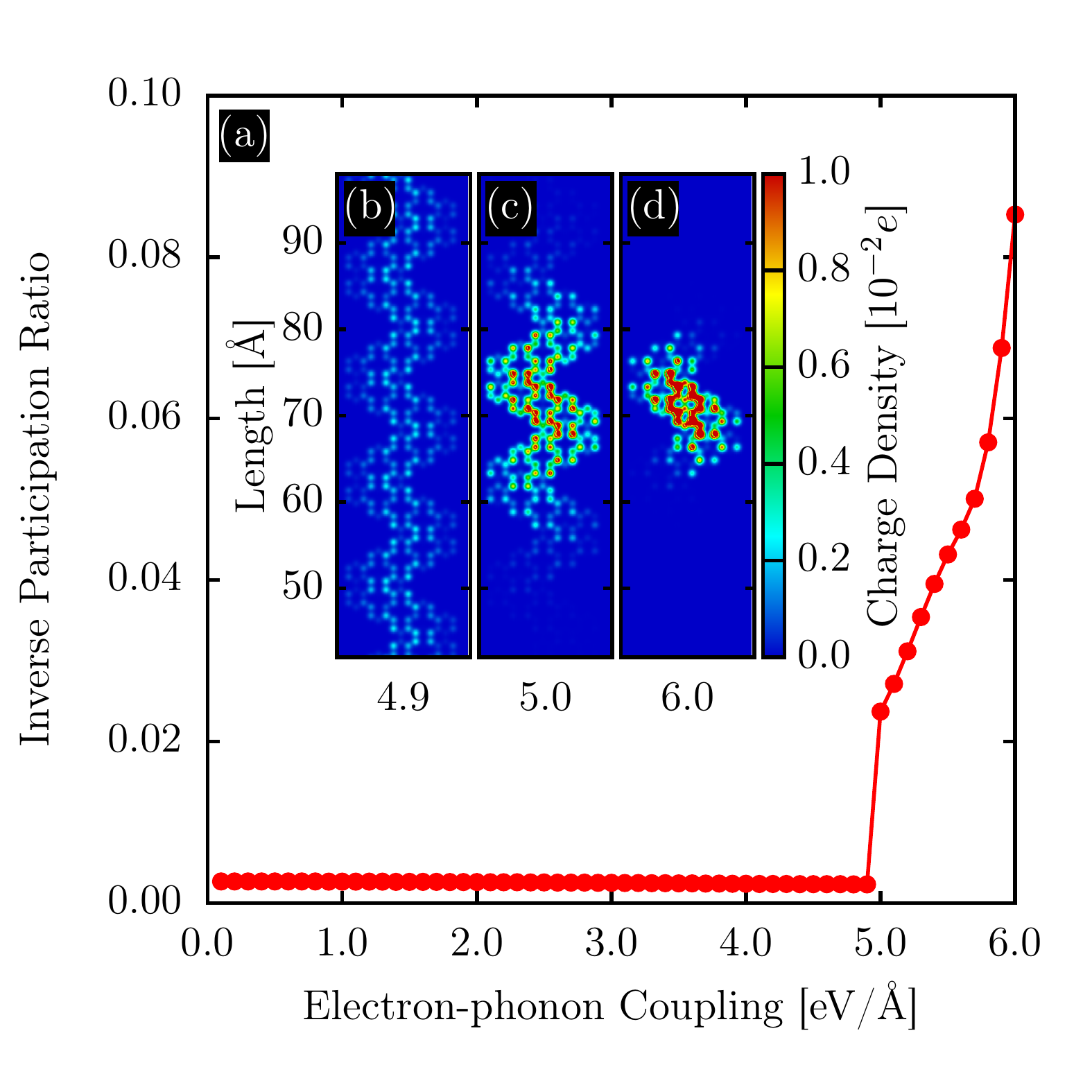}
	\caption{(a) Inverse Participation Ratio as a function of the electron-phonon coupling constant; (b), (c), and (d) charge distribution for small, critical, and large $\alpha$, respectively.}
	\label{fig4}
\end{figure}

\section{Results}
Here, we describe symmetrical shaped C--GNR (Figure \ref{fig1}). The total length of the unit cell is set to be 144 \AA, and periodic boundary conditions were applied in this direction. For the sake of clarity, we depict the C--GNRs as simple rectangular stripes, although the actual simulations were performed with the proper topology of the system shown above.

We begin our discussions by investigating the crucial aspect of the electron-phonon coupling constant value to be employed in our simulations. Our procedure is to tune the value of this constant having the energy gap of the system as a parameter. Indeed, the energy gap is the experimental property most convenient in determining the electron-phonon coupling value. The energy gap has to do with both the geometric disposition, which is a clear lattice property, and with the electronic distribution. The energy gap has also an important influence on electronic properties of the system which, as we shall see.

Although the energy gap of C--GNRS is still a subject of some debate in the literature, the accepted values lie around 2.28 eV \cite{chev}. Slight discrepancies of the energy gap can be attributed to impurities, the topology of the system, strain, and other external factors. Therefore, even though the reference value of 2.28 eV is a reliable one, other values --- as long as sufficiently close to this one --- should also be considered, as they might be representatives of the system under slightly different conditions. In this sense, we carried out a systematic spanning of the electron-phonon coupling values and performed simulations to calculate the HOMO-LUMO energy, the density of states and band structure, for each case. Figure \ref{fig3}(a) presents the result of this set of simulations. As $\alpha$ increases so do the energy gap of the system, which is an expected feature for low dimensional systems \cite{Hague_2011}. The figure highlights the $\alpha$ value corresponding to the expected 2.28 eV bandgap: the resulting electron-phonon coupling constant is 5.2 eV/\AA. Figure \ref{fig3}(b) represents the Density of States corresponding to $\alpha = 5.2 $ eV/\AA. Figure \ref{fig3} (c) presents a representation of the band structure of the system.  

\begin{figure}[]
	\centering
	\includegraphics[width = 0.7\linewidth]{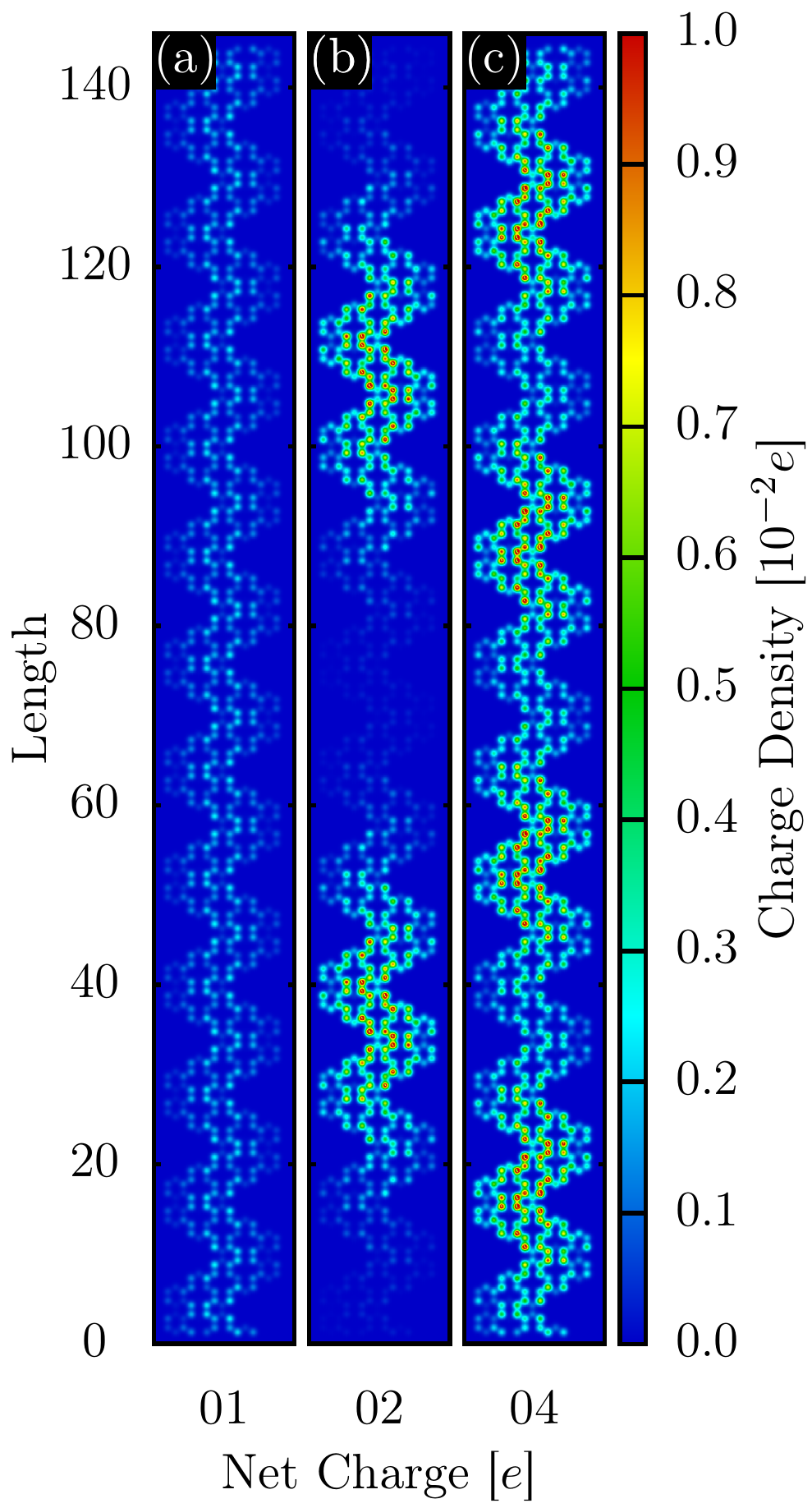}
	\caption{Atomic charge distribution for $\alpha = 4.5$ eV/\AA~in the case of one, two, and four holes in the system.}
	\label{fig5}
\end{figure}

\begin{figure*}[]
	\centering
	\includegraphics[width = 0.9\linewidth]{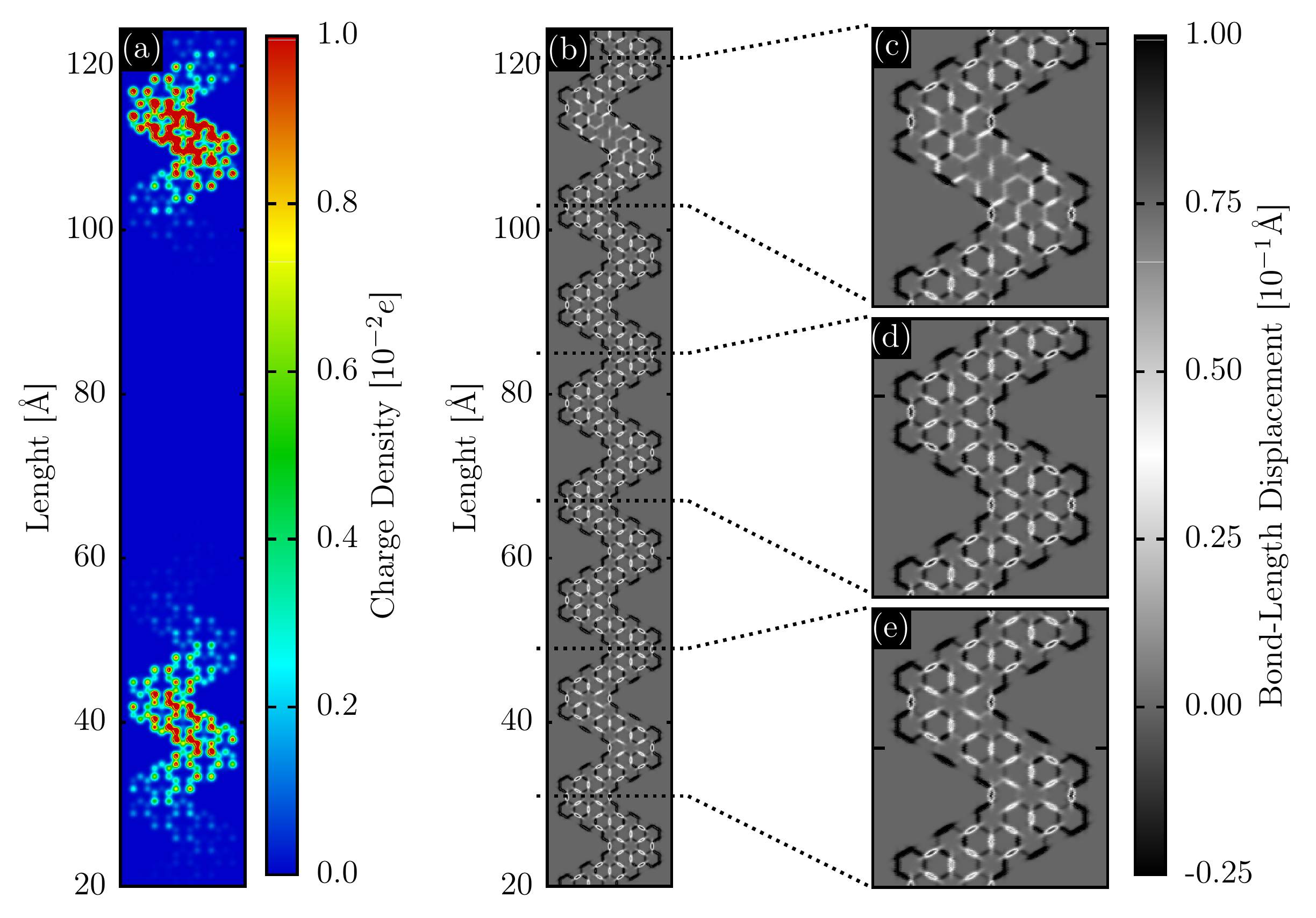}
	\caption{Ground sate (stationary) solution for a C--GNR lattice containing a polaron and a bipolaron. Panels (a) and (b) illustrate the charge density and bond-length profiles. Panels (c-e) zoomed in three distinct regions for the bond-length configuration panel: (c) the region containing a bipolaron, (d) a region without additional charge, and (d) the region containing a polaron. }
	\label{fig6}
\end{figure*}

Now that a suitable range for the electron-phonon constant ($\alpha$) is defined, we consider charge carriers in our simulations. A first step in doing so is to simulate the system with a net positive charge and to describe how the charge is distributed throughout the two-dimensional lattice. This is accomplished by extracting one electron from the C--GNR lattice and to compute the Inverse Participation Ratio (IPR), which is a measure of the degree of charge localization, commonly used to define the nature of quasiparticles\cite{holstein}:
\begin{equation}
	IPR = \frac{\sum_{i,j} |\rho_{i,j}|^2}{\left( \sum_{i,j} |\rho_{i,j}| \right)^2}.
\end{equation}
IPR is a dimensionless quantity that varies from 0.0 in the case of a completely
delocalized charge to 1.0 associated with complete localization. Figure \ref{fig4}(a) presents the result of the system's IPR for different $\alpha$ constants considered in this work and Figures \ref{fig4}(b-d) illustrate the atomic charge density profiles for C--GNR lattices considering representative electron-phonon coupling cases. In Figure \ref{fig4}(a) we have the behavior of the IPR as a function of $\alpha$. As is the case of many extended organic systems, our simulations presented relatively small values of IPR even for the cases where quasiparticles are present. This fact confirms the trend that suggests C--GNRs to be good conductors \cite{vovo}, as the more delocalized the charge carriers, the greater their mobilities. The figure is clear in indicating the critical value of 5.0 eV/\AA~as a phase transition between a delocalized picture and a quasiparticle mediated regime. Below this value, one has a completely delocalized scenario (Figure \ref{fig4}(b)). Above it, charge localization takes place in the form of quasiparticles, as can be inferred from Figures \ref{fig4}(c-d). 

Figures \ref{fig4}(b-d) present the profile of charge concentration for three different electron-phonon coupling constants, i.e., 4.9 (complete delocalization), 5.0 (critical value), and 6.0 eV/\AA~(localized charge). Simulations with $\alpha$ smaller than 4.9 eV/\AA~yield the same profile, as can be seen in (a), i.e., a null IPR is achieved in all cases. From 5.0 eV/\AA~on, the localization degree monotonically increases and a quasiparticle scenario is always observed. The high value of 6.0 eV/\AA~is associated with a correspondingly high IPR. This should have a large impact on the charge mobility of the particular system. However, one should be most interested in values slightly higher than 5.0 eV/\AA, as this constant corresponds to the experimentally observed bandgap value.  

\begin{figure*}[]
	\centering
	\includegraphics[width = \linewidth]{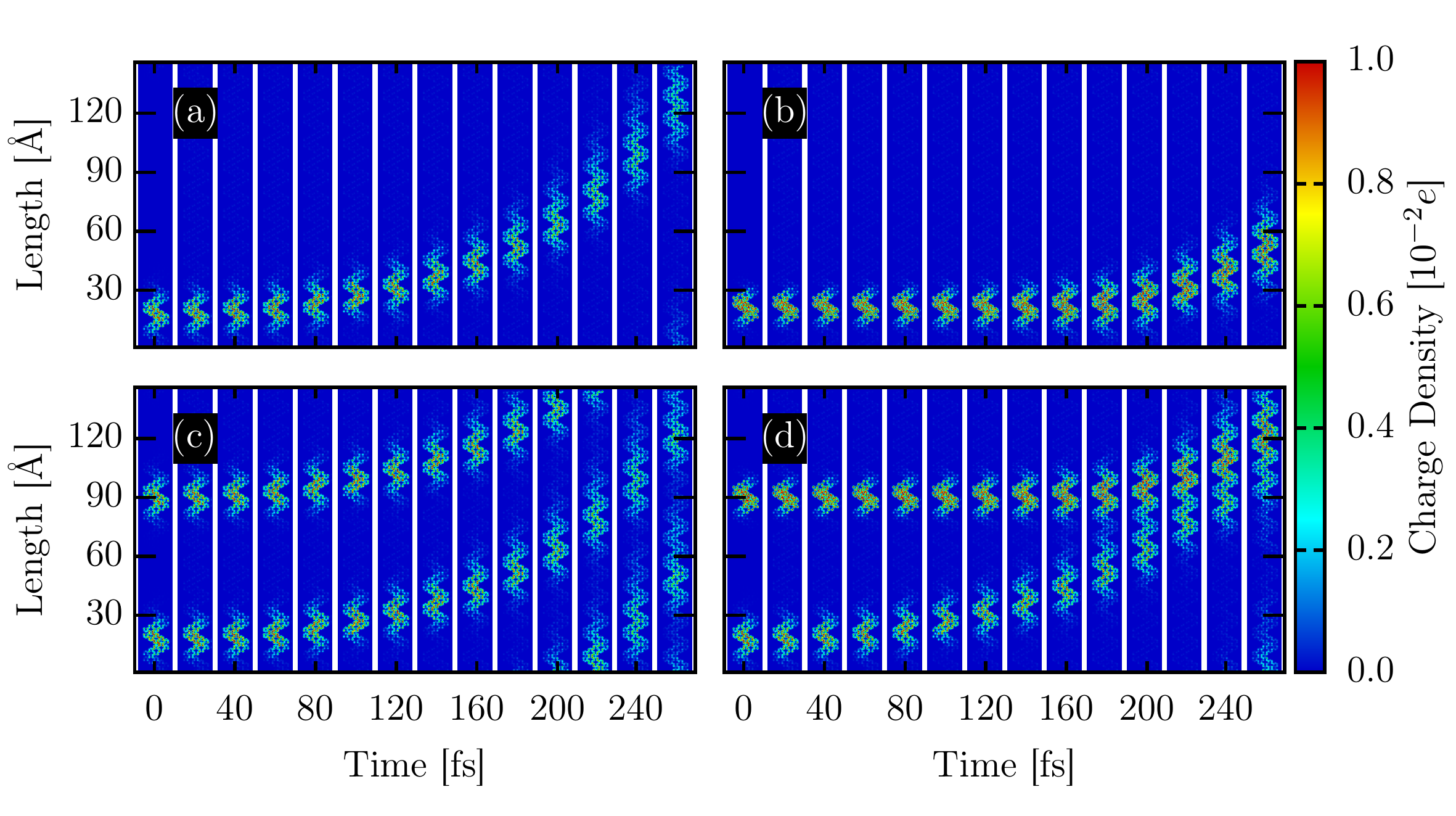}
	\caption{Dynamics of (a) Polaron, (b) Bipolaron, (c) pair of Polarons, and (d) Polaron and Bipolaron on Chevron-graphene nanoribbons under the influence of a 1.0 mV/\AA~ electric field applied in the length direction.}
	\label{fig7}
\end{figure*}

A natural question that arises is whether the same behavior is observed when a more charged system is considered. To investigate this trend, we consider an electron-phonon coupling regime smaller than the critical value. Figure \ref{fig5} corresponds to C--GNRs in which $\alpha = 4.5 $ eV/\AA~is considered. In Figure \ref{fig5}(a) we have one positive charge carrier in the system. As this value is below the 5.0 eV/\AA~threshold, the result is of a completely delocalized picture, indistinguishable from that of Figure \ref{fig4}(b). The situation is drastically altered when two charge carriers are considered, as in Figure \ref{fig5}(b). Note that, in this case, even though $\alpha < 5.0 $ eV/\AA~a clear charge localization in two centers is observed. In Figure \ref{fig5}(c), we include four holes in the system and, again, one can see the charge distribution typical of quasiparticles. In this case, because of the size of the nanoribbons, the four quasiparticles cover the whole extension of its length. The periodic disposition of variable amounts of charge, however, leaves no room for doubt: far from being a delocalized evenly displaced scenario as in Figure \ref{fig5}(a), we have four equally spaced charge carriers. From this set of simulations, we conclude that the critical value that separates delocalized and localized scenarios is highly dependent on the amount of charge. Highly charged C--GNRs tend to favor a charge carrier-mediated transport over a delocalized picture. 

We now focus our attention on the experimentally expected value of $\alpha = 5.2$ eV/\AA. The idea here is to study the system with different quasiparticle scenarios. Figure \ref{fig6}(a) presents the charge density localization of C--GNR with a polaron (charge localization signature between 20-60~\AA) and a bipolaron (charge localization signature between 100-120~\AA). Figures \ref{fig6}(b-e), in turn, illustrate the respective bond-length configuration. For the sake of clarity, Figures \ref{fig6}(c-e) zoomed in three distinct regions for the bond-length configuration panel: (c) the region containing a bipolaron, (d) a region without additional charge, and (d) the region containing a polaron. By keeping a $+2e$ net charge but considering the opposite spin we achieve the hole bipolaron in Figure \ref{fig6}(a). Note the stronger red pattern when compared to the charge density for a bipolaron with a polaron, which is associated with twice the charge of the individual carrier of the latter. Importantly, in Figures \ref{fig6}(b-e) one can realize that the charge localization signatures for a polaron and a bipolaron have related local lattice distortions. This behavior is a piece of evidence that two different structures formed by a self-interacting state between extra charge and local lattice deformations. Moreover, one can note that the lattice distortions associated with the bipolaron present a different pattern when it comes to the polaron case. This feature is attributed to the higher degree of charge localization as presented for the bipolaron case.  

\begin{figure}[]
	\centering
	\includegraphics[width = \linewidth]{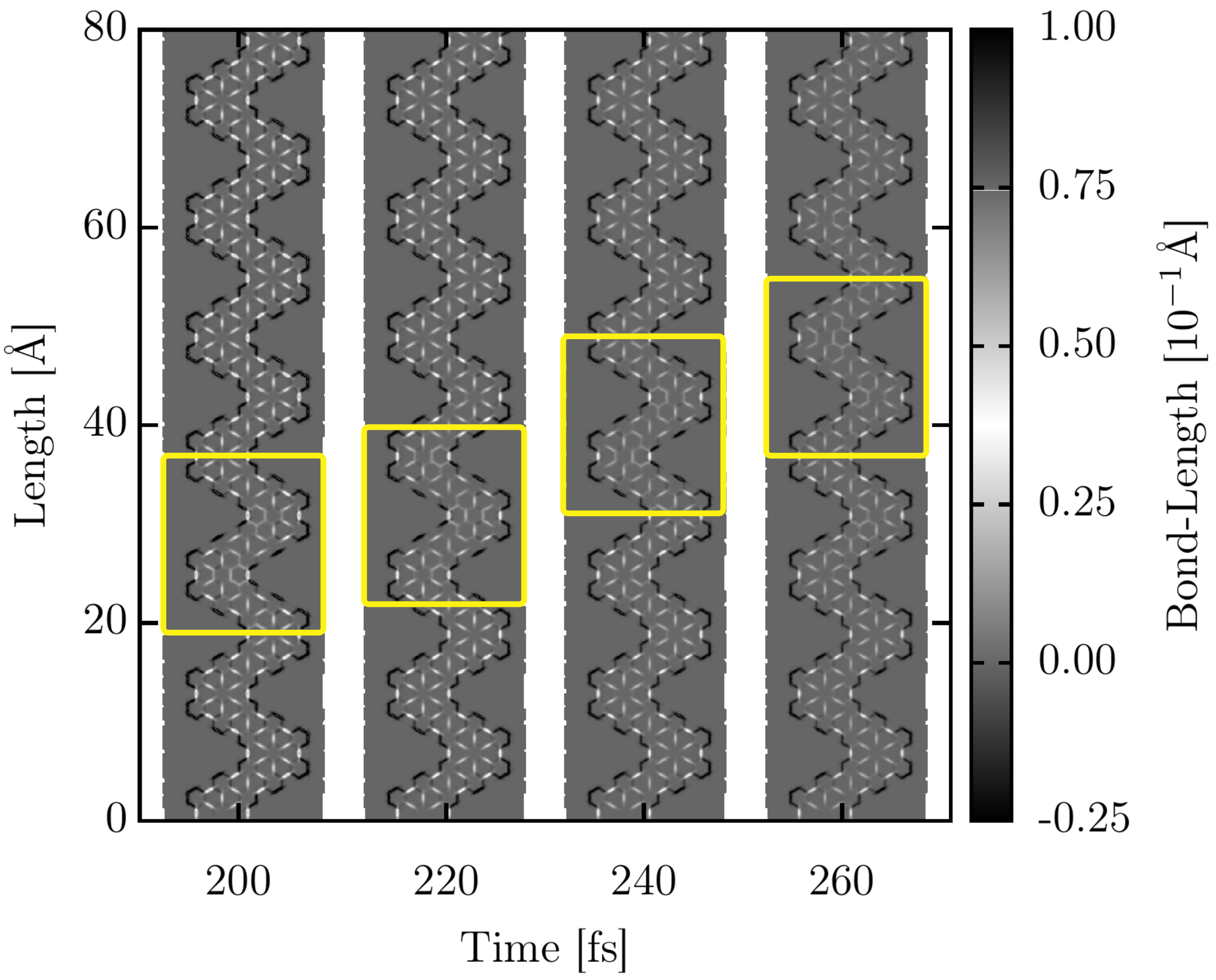}
	\caption{Time evolution for the bond-length configuration of the bipolaron dynamics presented in Figure \ref{fig7}(b). The yellow rectangles serve as a guide for the eye in following the bipolaron's bond-length displacements.}
	\label{fig8}
\end{figure}

\begin{figure*}[]
	\centering
	\includegraphics[width = \linewidth]{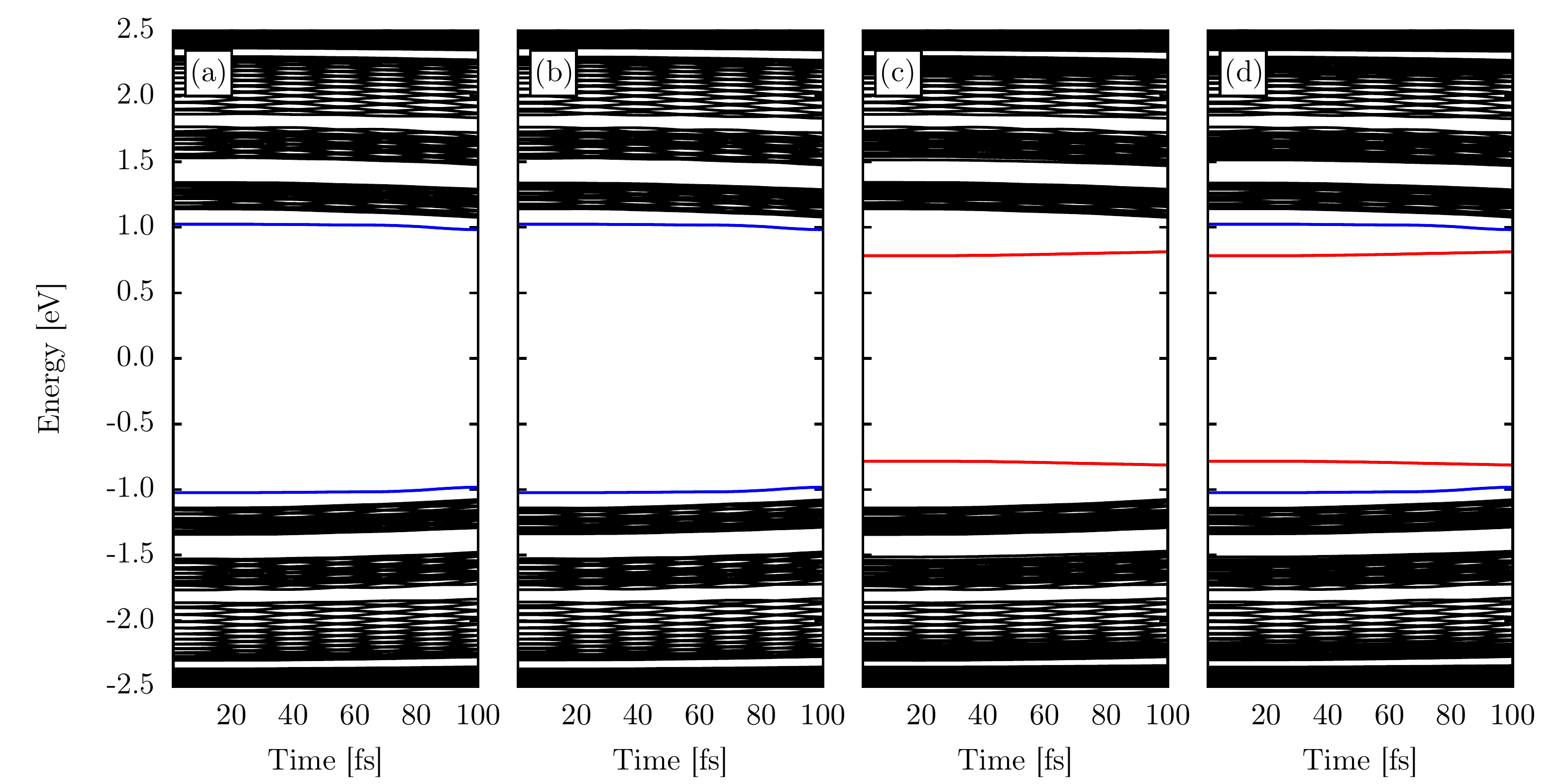}
	\caption{Time evolution profile of the intra-gap energy levels of simulations shown in Figure \ref{fig7}.}
	\label{fig9}
\end{figure*}

The dynamics of the charge carriers is presented in Figure \ref{fig7}. This analysis is crucial on defining the mobility and dynamical stability of the carriers which, in turn, impact in transport properties of the system. We investigate how each system behaves under the influence of a 1.0 mV/\AA~electric field. Figure \ref{fig7}(a) presents the polaron dynamics. Note that, after a transient period the polaron begins to move as a result of the electric field. It should be remembered that the electric field is turned on adiabatically according to the approach in the reference \cite{cunha_PRB}, thus the delay in the polaron response. In Figure \ref{fig7}(b) we observe the dynamics of the bipolaron. More massive, its response is accordingly slower and, although being more charged, it results in a shorter displacement. Figure \ref{fig7}(c) presents two polarons simulation. As the quasiparticles are identical, their path are parallel. Each one is similar to Figure \ref{fig7}(a). In this case, note the periodic boundary conditions applied to the system: as the upper polaron reaches the top of the figure, it comes right back from its bottom. This does not happen for the polaron in the final Figure \ref{fig7}(d). In this case, we present the simulation in which both a polaron and a bipolaron are present. The aforementioned difference of inertia and the consequent difference in response to the applied field becomes clear, as the bipolaron barely moves 30 \AA~while, at the same time, the same electric field makes the polaron to displace more than 80 \AA. Also, note that a collision takes place between the faster polaron and the slower bipolaron, which prevents the former to achieving the edge of the nanoribbon, as in Figure \ref{fig7}(c).

As mentioned above, polarons and bipolarons move in the lattice as a composited state formed by the self-interaction between the additional charge and related local lattice distortions. To show this feature, Figure \ref{fig8} presents the bipolaron dynamics corresponding to Figure \ref{fig7}(b). The yellow rectangles serve as a guide for the eye in following the bipolaron's bond-length displacements. In this way, in Figure \ref{fig8} one can note that the local lattice distortions move by following the charge density motion. This dynamical trend is similar for the other cases discussed here (Figures \ref{fig7}(a,c,d)). Therefore, the combined analyses of Figures \ref{fig7} and \ref{fig8} states that in C--GNRs the composite quasiparticles are dynamically stable, once charge and related lattice deformations move together as a single structure. 

As can be inferred from Figure \ref{fig7}, after a transient period for acceleration, both charge carriers reach their saturation velocities. These terminal velocities are 0.4~\AA/fs and 0.12~\AA/fs for the polaron and bipolaron cases, respectively. When it comes to standard armchair graphene nanoribbons, the terminal velocities of polarons and bipolarons can range within the intervals 0.3-5.1 \AA/fs \cite{gesiel_PCCP} and 0.05-1.30 \AA/fs \cite{silva_SR}, respectively, depending on the ribbon's width and the strength of the applied electric field. Importantly, the similarities in terminal velocities among C--GNRs and standard armchair graphene nanoribbons comes from their comparable charge localization profiles that yield quasiparticles with equivalent effective masses. 

Finally, we present in Figure \ref{fig9} the time evolution profile of the intra-gap energy levels of the simulations shown in Figure \ref{fig7}. For the sake of convenience, this figure illustrates only the first 100 fs of the carriers' dynamics. The blue and red intra-gap levels are referring to the polaron and bipolaron species, respectively. In Figures \ref{fig9}(a-d) one can note that the quasiparticle levels remain within the bandgap during the dynamics. This behavior is one more piece of evidence suggesting that polarons and bipolarons are dynamically stable in C--GNRs. The bipolaron levels are positioned narrower within the bandgap indicating that bipolarons are more stable species than polarons. Such a trend for the quasiparticles stability was also found for standard armchair graphene nanoribbons \cite{silva_SR}.  

\section{Conclusions}
We have carried out an extensive investigation of the static and dynamical properties of different quasiparticles in Chevron-graphene nanoribbons. Our computational protocol is based on a 2D tight-biding Hamiltonian that includes lattice relaxation effects. The reported results were observed to be rather different from those described for standard graphene nanoribbons, which may pave the way for different applications. The influence of the electron-phonon coupling on the conduction mechanism was explored. Here, we have obtained 5.2 eV/\AA~as the electron-phonon coupling strength corresponding to the expected 2.28 eV bandgap for C--GNRs. This value is close to the critical value that determines the phase transition between a delocalized charge and a quasiparticle transport. We characterized and performed a detailed analysis of the dynamics of different charge carriers (polarons and bipolarons). By means of these results, we were able to determine the kind of quasiparticle presenting higher mobility. The terminal velocities for polarons and bipolarons in C--GNR lattices are 0.4~\AA/fs and 0.12~\AA/fs, respectively. These velocities are similar to what is found for standard armchair graphene nanoribbons. The time evolution for the intra-gap energy levels suggest that polarons and bipolarons are dynamically stable in C--GNRs. Particularly, the bipolaron levels are positioned narrower within the bandgap indicating that this species is more stable than polarons. 

\printcredits

\bibliographystyle{cas-model2-names}

\bibliography{cas-refs}

\end{document}